\documentclass[twocolumn,aps,showpacs,prb,tightenlines,amsmath,amssymb]{revtex4-1}

\usepackage{graphicx}
\usepackage{amssymb}               
\usepackage{amsmath}
\usepackage{colordvi} 

\begin{document} 

\title{Effect of nonequilibrium phonons on hot-electron spin relaxation in
$n$-type GaAs quantum wells}
 
\author{P. Zhang}
\author{M. W. Wu}
\thanks{Author to whom correspondence should be addressed}
\email{mwwu@ustc.edu.cn.}
\affiliation{Hefei National Laboratory for Physical Sciences at
  Microscale and Department of Physics, 
University of Science and Technology of China, Hefei,
  Anhui, 230026, China}
\date{\today}

\begin{abstract} 
We have studied the effect of nonequilibrium longitudinal optical phonons on
hot-electron spin relaxation in $n$-type GaAs quantum wells. The longitudinal
optical phonons, due to the finite relaxation rate, are driven to 
nonequilibrium states by electrons under an in-plane electric
field. The nonequilibrium phonons then in turn influence the electron spin
relaxation properties via modifying the electron heating and
drifting. The spin relaxation time is elongated due to the enhanced
electron heating and thus the electron-phonon scattering in the presence of nonequilibrium phonons. The frequency of spin precession, which is
roughly proportional to the electron drift velocity, can be either
increased (at low electric field and/or high lattice temperature) or
decreased (at high electric field and/or low lattice temperature). 
The nonequilibrium phonon effect is more pronounced when 
the electron density is high and the impurity density is low. 
\end{abstract}

\maketitle

\section{Introduction}
Understanding spin relaxation is an important issue for the possible application of spintronic 
devices.\cite{aws,zutic,dy,wu-rev} Among different kinds of spin relaxation
mechanisms,\cite{dp,bap,ey} scattering plays an essential
role. In general cases, phonons are assumed to form 
an equilibrium bath when carrier-phonon
scattering is considered. This treatment works well when
the carrier system is near the equilibrium. If the carriers are far away from the equilibrium
(e.g., driven by an electric field or excited by a laser beam),
phonons can be driven to run away from their 
equilibrium states significantly by carriers when the carrier 
energy relaxation mainly
goes through the phonon emissions and the phonon relaxation time is
comparable with (or longer than) the carrier-phonon scattering
time. The nonequilibrium phonons in turn are able to affect the
electron dynamics, including the spin relaxation. 
In fact, the hot-electron transport with
nonequilibrium phonons has been
investigated,\cite{ramonas,mick193312,cai2636,lei6281,cai8573,vai9886,vai13082,pau5580,wu5438}
showing that the calculated electron energy loss rate and mobility
fit better with experimental data than those obtained with the 
equilibrium phonons.\cite{lei6281,cai2636,cai8573} These studies also  
indicate that it is necessary to treat phonons as nonequilibrium ones in
the hot-carrier system, and the nonequilibrium phonons may
affect spin relaxation via modifying the carrier heating and
drifting. 

The hot-electron spin relaxation/dephasing has been studied
theoretically in both (001) quantum-well
structures\cite{weng1,weng2,zhang,zhou} and bulk materials,\cite{jiang} by means of the kinetic spin Bloch equation
(KSBE) approach.\cite{wu-rev} The spin relaxation/dephasing time is
found to increase with electric field when both the temperature and
electric field are low, especially in high mobility
samples.\cite{weng1,weng2,zhang,zhou,jiang} When the electric field is
high [for which the multi-subband (in confined nanostructures)\cite{weng2} and/or
multi-valley\cite{zhang} effect have to be taken into account], the spin
relaxation/dephasing time decreases with electric field.\cite{weng1,weng2,zhang,jiang} In these studies the phonons are treated as
equilibrium ones. This work is to investigate the influence 
of nonequilibrium phonons on hot-electron
spin relaxation in an $n$-type GaAs quantum well, where the spin-orbit
coupling term is the Dresselhaus type\cite{dresselhaus,zhang} and the
spin relaxation is limited by the D'yakonov-Perel' mechanism.\cite{dp}
 
The paper is organized as follows. In Sec.~II we set up the model and the
KSBEs with nonequilibrium phonons. In Sec.~III the
effect of nonequilibrium phonons on spin relaxation is
investigated. Finally, we conclude in Sec.~IV.

\section{Model and KSBEs}
We start our investigation from an $n$-type $(001)||{\bf \hat{z}}$ GaAs quantum well with
an in-plane electric field. The well width $a=5$~nm. Only the lowest
subband is relevant with the proper 
electron density $N_e$, lattice temperature $T_L$ and electric field
${\bf E}$. Due to the electron
localization in the ${\bf\hat{z}}$-direction, the electron-phonon coupling is
spatially inhomogeneous, i.e., the emission and absorption of phonons
mainly occur in the well where electrons have substantial
density. If the phonon relaxation is fast enough or the phonons
[particularly, the acoustic (AC) phonons] can easily
penetrate through the well interfaces, these phonons can be deemed as in
equilibrium with the bulk modes. In our study we assume that the 
AC phonons keep in 
equilibrium and the longitudinal-optical (LO) phonons are
nonequilibrium.\cite{lei6281,cai2636,mick193312,pau5580,vai9886,vai13082}
In order to investigate the spin relaxation of electrons which are
inhomogeneously coupled with the nonequilibrium LO phonons, 
we combine the rate equation of the LO phonons
[Eq.~(\ref{ksbelo})], described as
``quasi-2D'',\cite{lei6281,cai2636,cai8573} with the electron KSBEs [Eq.~(\ref{ksbee})]:\cite{wu-rev}
\begin{eqnarray}
    \frac{\partial \rho_{\bf k}}{\partial t}&=&\left.\frac{\partial\rho_{\bf k}}{\partial t}\right|_{\rm dri}+\left.\frac{\partial\rho_{\bf k}}{\partial t}\right|_{\rm coh}
  +\left.\frac{\partial\rho_{\bf k}}{\partial t}\right|_{\rm
    scat},\label{ksbee}\\
  \frac{\partial n_{\bf q}}{\partial t}&=&\left.\frac{\partial n_{\bf q}}{\partial t}\right|_{\rm
    scat}.\label{ksbelo}
\end{eqnarray}
In Eq.~(\ref{ksbee}), $\rho_{\bf k}$ represent the
density matrices of electrons with in-plane momentum ${\bf k}$, whose diagonal terms
$\rho_{{\bf k}\sigma\sigma}\equiv f_{{\bf
    k}\sigma}$ ($\sigma=\pm1/2$) represent the electron
distribution functions and the off-diagonal ones $\rho_{{\bf
    k}\frac{1}{2}-\frac{1}{2}}=\rho_{{\bf  k}-\frac{1}{2}\frac{1}{2}}^\ast$
 describe the inter-spin-band correlations for the spin
 coherence. $\left.\frac{\partial \rho_{\bf
    k}}{\partial t}\right|_{\rm dri}=e{\bf
  E}\cdot\mbox{\boldmath$\nabla$\unboldmath}_{{\bf
    k}}{\rho}_{{\bf k}}$
are the driving terms from the external electric field.
$\left.\frac{\partial \rho_{\bf
    k}}{\partial t}\right|_{\rm coh}$ are the coherent terms describing the 
coherent spin precessions due to the effective magnetic
fields from the Dresselhaus term\cite{dresselhaus,zhang} and the
Hartree-Fock Coulomb interaction, as well as the optional 
external magnetic field in the Voigt configuration.
$\left.\frac{\partial \rho_{\bf k}}{\partial t}\right|_{\rm scat}$
stand for the scattering terms of electrons, including the electron-LO/AC
phonon, electron-impurity
 and electron-electron Coulomb
 scatterings. $n_{\bf q}$ in Eq.~(\ref{ksbelo}) are the distributions of quasi-2D LO phonons with
 in-plane momentum ${\bf q}$. $\left.\frac{\partial n_{\bf
     q}}{\partial t}\right|_{\rm scat}$ stand for the scattering terms
of the LO  phonons, including the phonon-phonon and phonon-electron
 scatterings. Expressions of the coherent and scattering terms of electrons are given in
 detail in Refs.~\onlinecite{weng1} and \onlinecite{zhou}, except that the electron-LO phonon
 scattering term should be slightly modified here as the LO phonons are described as
 quasi-2D (this modification makes no difference when the LO
 phonons are in equilibrium).\cite{lei6281,cai2636,cai8573} The electron-LO phonon scattering term
 in Eq.~(\ref{ksbee}) reads
 \begin{eqnarray}\nonumber
\left.\frac{\partial \rho_{{\bf k}}}{\partial t}\right|_{\rm
  scat}^{e\mbox{-}{\rm LO}}&=&-\{S_{\bf k}(>,<)
    -S_{\bf k}(<,>)
    +S_{\bf k}(>,<)^{\dagger}\\ &&-S_{\bf k}(<,>)^{\dagger}\},
  \end{eqnarray}
  with
 \begin{eqnarray}\nonumber
    S_{\bf k}(>,<)&=&\frac{\pi}{A}\sum_{{\bf
        k}^\prime}M_{{\bf k}-{\bf k}^\prime}^2\rho_{{\bf
        k}^{\prime}}^{>}\rho_{{\bf
        k}}^{<} [n_{{\bf k^\prime}-{\bf k}}^{<}
    \delta(\varepsilon_{{\bf k}}-\varepsilon_{{\bf
        k}^{\prime}}+\hbar\omega_0)
    \\ &&+n_{{\bf k}-{\bf k^\prime}}^{>}\delta(\varepsilon_{{\bf k}}
    -\varepsilon_{{\bf  k}^{\prime}}-\hbar\omega_0)]
  \end{eqnarray}
 and $S_{\bf k}(<,>)$ obtained by interchanging $>$ and $<$ from
   $S_{\bf k}(>,<)$. The scattering term in Eq.~(\ref{ksbelo}) reads
 \begin{eqnarray}\nonumber
    \left.\frac{\partial n_{{\bf q}}}{\partial t}\right|_{\rm
      scat}&=&-\frac{n_{{\bf q}}-n^0_{\bf
        q}}{\tau_{pp}}-\frac{2\pi}{A} M_{\bf q}^2\sum_{\bf k}\delta(\varepsilon_{\bf k}-\varepsilon_{\bf
      k-q}-\hbar\omega_0)\\ && \times[{\rm
      Tr}(\rho_{\bf k}^>\rho_{\bf k-q}^<)n_{{\bf
        q}}^<-{\rm Tr}(\rho_{\bf k}^<\rho_{\bf k-q}^>)n_{{\bf
        q}}^>].
\label{phonon}
  \end{eqnarray}
 In the above equations $\rho_{\bf k}^{<}=\rho_{\bf k}$, $\rho_{\bf
    k}^{>}=1-\rho_{\bf k}$, $n_{\bf q}^{<}=n_{\bf q}$ and $n^>_{\bf q}=n_{\bf
  q}+1$. $\varepsilon_{\bf
    k}=\hbar^{2}{\bf k}^{2}/2m^{*}$ represents the energy of electron with momentum
  ${\bf k}$ and effective mass $m^{*}=0.067m_0$, and
  $\hbar\omega_0=35.4$~meV is the energy of the LO phonons.\cite{lei9134} $A$ is the
  area of the well layer. $M_{\bf q}^2=\frac{1}{L}\sum_{q_z}g_{qq_z}^2|I(iq_z)|^2$ is
  the effective electron-LO phonon scattering matrix element with
  $g_{qq_z}^2=\frac{e^{2}\omega_0}{2\hbar^2\epsilon_{0}(q^{2}+q_{z}^{2})}(\kappa_{\infty}^{-1}-\kappa_{0}^{-1})$.\cite{lei9134} 
  $L$ is the size of the sample along the ${\bf \hat{z}}$-direction. $\kappa_{0}=12.9$ and $\kappa_{\infty}=10.8$ are the
  relative static and high-frequency dielectric constants
  respectively,\cite{lei9134} and $\epsilon_{0}$ is the vacuum dielectric constant. 
  $|I(iq_{z})|^{2}=\frac{\pi^{4}\sin^{2}y}{y^{2}(y^{2}-\pi^{2})^{2}}$
  with $y=\frac{aq_z}{2}$ stands for the form factor under the infinite-depth well approximation. The first term on the right hand side of
  Eq.~(\ref{phonon}) represents the contribution
  from the phonon-phonon scattering in relaxation time
  approximation. $n_{\bf q}^0=[\exp(\hbar\omega_0/k_BT_L)-1]^{-1}$ is the number 
  of quasi-2D LO phonons in equilibrium with the 
AC phonons. The population relaxation time $\tau_{pp}$ is
  contributed by anharmonic lattice vibrations (especially the
  third-order anharmonicity), which in principle depends on the
  phonon momentum and lattice temperature. Moreover, distinctly heated
  nonequilibrium LO phonons (depending on the heating and relaxation
  of electrons) may have different relaxation times. In spite of these
  intricate factors involved in the LO phonon relaxation, we assume
  $\tau_{pp}$ to be a constant only depending on the lattice
  temperature, by adopting $\tau_{pp}$ fitted by  Vall\'ee and Bogani from the time-resolved
  coherent anti-Stokes Raman scattering experiment.\cite{vallee12049} It gives
  $\tau_{pp}(T_L)=\tau_{pp}^0/\{1+[\exp(0.2\hbar\omega_0/k_BT_L)-1]^{-1}+[\exp(0.8\hbar\omega_0/k_BT_L)-1]^{-1}\}$
  with $\tau_{pp}^0\equiv\tau_{pp}(0)\approx 9$~ps.\cite{vallee12049} This formula in
  fact depicts the dominant decay route of an LO phonon near the center of
  the Brillouin zone into a transverse
  AC phonon and a different LO phonon at the $L$ critical point of the
  Brillouin zone. The relaxation time approximation with a constant
  $\tau_{pp}$ related only to the lattice temperature has been widely utilized in the
  study of hot-electron transport with the presence of nonequilibrium LO
  phonons.\cite{lei6281,cai2636,cai8573,vai13082,vai9886}
 
\section{Results}

We numerically solve the KSBEs following the scheme mainly laid out in Ref.~\onlinecite{weng1},
with the rate equation of the LO phonons discreted in the momentum
space in a way similar to that for electrons. The impurity
density is set as zero and the electric field ${\bf E}=-E{\bf \hat{x}}$
with $E\ge 0$. No magnetic field is applied except otherwise
specified. The initial conditions at time $t=0$ are chosen 
as the steady-state solution
 of Eqs.~(\ref{ksbee}) and (\ref{ksbelo}) in the absence of the
spin-orbit coupling in the coherent term $\left.\frac{\partial \rho_{\bf k}}{\partial
    t}\right|_{\rm coh}$.\cite{weng1} Numerically, they are prepared from a state at $t=-t_0$ ($t_0>0$) with $f_{{\bf k}\sigma}(-t_0)=\{\exp[(\hbar^2{\bf
  k}^2/2m^\ast-\mu_\sigma)/k_BT_L]+1\}^{-1}$, $\rho_{{\bf
    k}\sigma-\sigma}(-t_0)=0$ and $n_{\bf q}(-t_0)=n_{\bf 
  q}^0$. Here $\mu_\sigma$ are the electron chemical potentials determined
by $\frac{1}{A}\sum_{\bf k}\mbox{Tr}[\rho_{\bf  k}(-t_0)]=N_e$ and
$\frac{1}{A}\sum_{\bf k}\mbox{Tr}[\rho_{\bf  k}(-t_0)\sigma_z]=N_eP_0$,
where $P_0=0.05$ is the spin polarization and
$N_e=4\times 10^{11}$~cm$^{-2}$ is the electron area density. With the
driving from the electric field and the scattering, the system reaches a
steady state at time $t=0$. After time $t=0$, the spin-orbit
  coupling in the coherent term is switched on and electron spins begin to relax with an initial
spin polarization $P(t=0)=P_0=0.05$. The spin relaxation time $\tau$
is obtained from the time evolution of spin 
polarization $P(t)=\frac{1}{AN_e}\sum_{\bf k}\mbox{Tr}[\rho_{\bf
  k}(t)\sigma_z]$, the electron drift velocity is the steady value of ${\bf
  v}(t)=\frac{1}{A}\sum_{\bf k}\mbox{Tr}[\rho_{\bf k}(t)]\hbar{\bf 
  k}/m^\ast=v_{x}(t){\bf \hat{x}}$ and the hot-electron temperature $T_e$
is fitted out from the Boltzmann tail of the steady-state electron
distribution.\cite{weng1,zhang}

\begin{figure}[ht]
  \hspace{0 cm}\includegraphics[width=9cm]{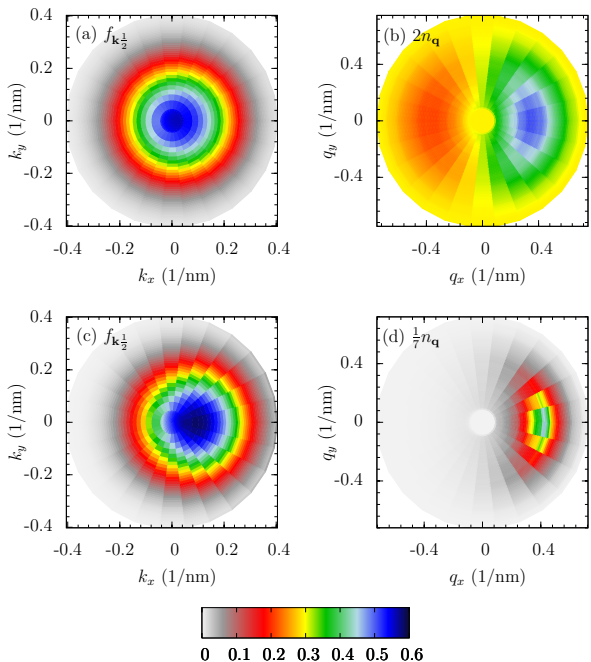}
  \caption{(Color online) Typical steady-state distributions of
    nonequilibrium electrons [(a) and (c)] and LO phonons [(b) and (d)] 
    in momentum space. (a) and (b): $T=200$~K and $E=0.3$~kV/cm; (c)
    and (d): $T=50$~K and $E=1$~kV/cm. Note that in (b) and (d)
    the LO phonon distributions $n_{\bf 
      q}$ are rescaled by a factor 2 and $\frac{1}{7}$, respectively.}    
\label{figzw1}
 \end{figure}

In Fig.~\ref{figzw1} the typical steady-state distributions of the
nonequilibrium electrons and LO phonons in momentum space are
plotted. It is shown by Fig.~\ref{figzw1}(a) and (c) that under the electric 
field along the $-{\bf \hat{x}}$-direction, the electrons gain a drift
velocity along the ${\bf
  \hat{x}}$-direction. Figure~\ref{figzw1}(b) and (d)
  show the corresponding distributions of nonequilibrium LO phonons. The LO
  phonons with either very large or small momenta [e.g., in the edge or center of the 
momentum space shown in Fig.~\ref{figzw1}(b) and (d)] are
in equilibrium with the AC phonons. It is seen from the figure that the equilibrium
distributions of the LO phonons are different due to the distinct lattice
temperatures [200~K for Fig.~\ref{figzw1}(b) and 50~K for
Fig.~\ref{figzw1}(d)]. It is interesting to see that the LO phonons with
mediate momenta are driven far away from their equilibrium states
by the hot-electrons. In the case with  
$T_L=200$~K and $E=0.3$~kV/cm [Fig.~\ref{figzw1}(b)], the LO phonons
with $q_x>0$ are emitted (and form a peak in the 
$q_x$-positive momentum region) and those with $q_x<0$ are
absorbed (and form a valley in the $q_x$-negative momentum
region). With the decrease of $T_L$ and/or the increase of $E$, the
valley in the $q_x$-negative momentum region is suppressed or even disappears, as shown in
Fig.~\ref{figzw1}(d) for the case with 
$T_L=50$~K and $E=1$~kV/cm. In any case, the total phonon density increases and a net 
positive momentum is gained by the LO phonons.

    \begin{figure}[ht]
      \begin{minipage}[]{10cm}
        \hspace{-1.5 cm}\parbox[t]{5cm}{
          \includegraphics[width=4.5cm,height=4.5 cm]{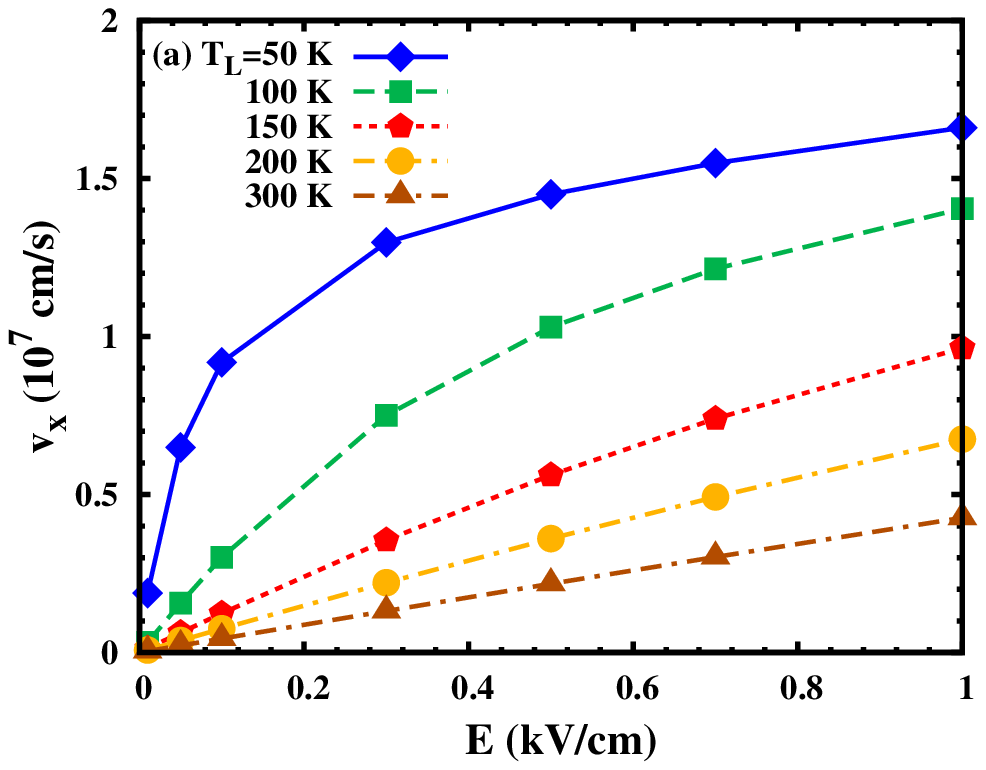}}
        \hspace{-0.7 cm}\parbox[t]{5cm}{
          \includegraphics[width=4.5cm,height=4.5 cm]{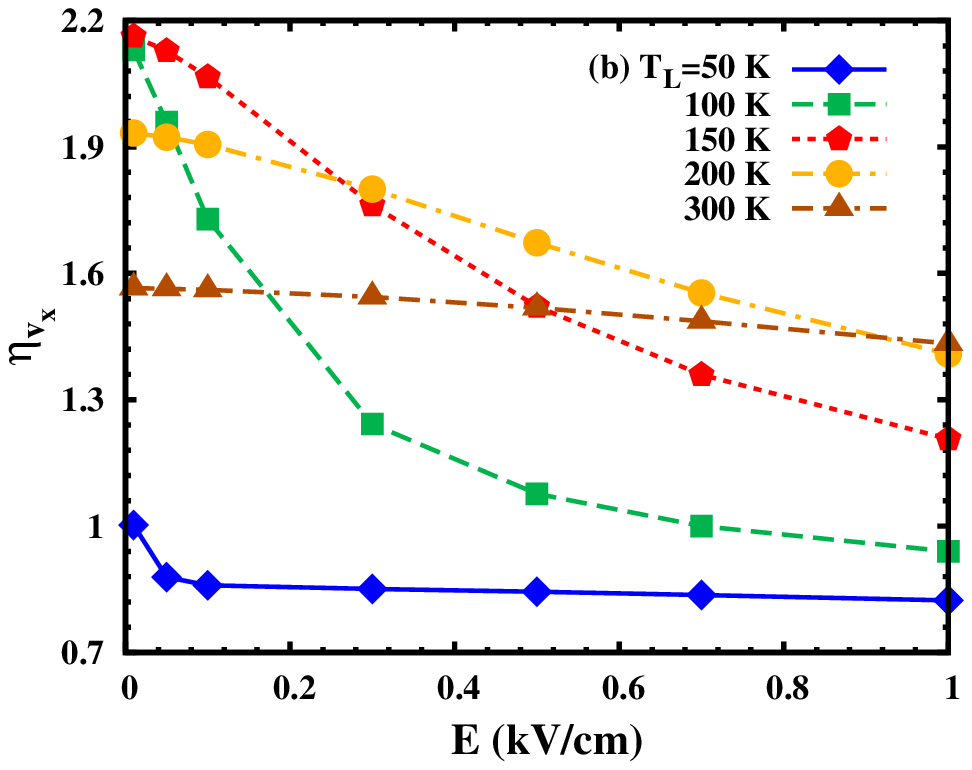}}
      \end{minipage}
      \begin{minipage}[]{10cm}
        \hspace{-1.5 cm}\parbox[t]{5cm}{
          \includegraphics[width=4.5cm,height=4.5 cm]{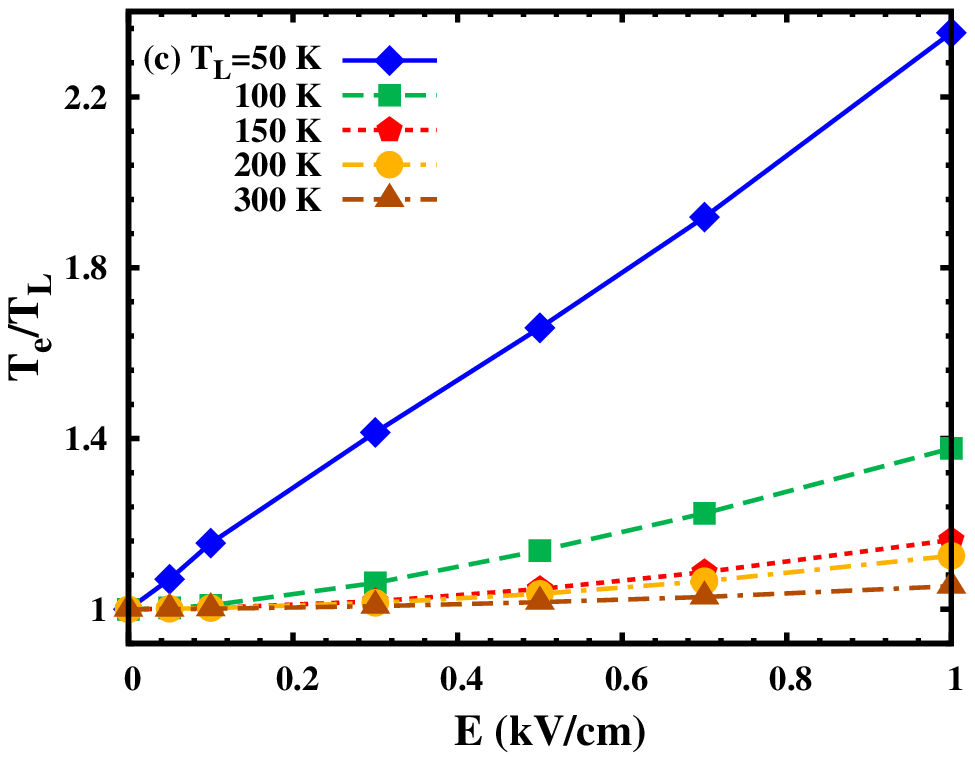}}
        \hspace{-0.7 cm}\parbox[t]{5cm}{
          \includegraphics[width=4.5cm,height=4.5 cm]{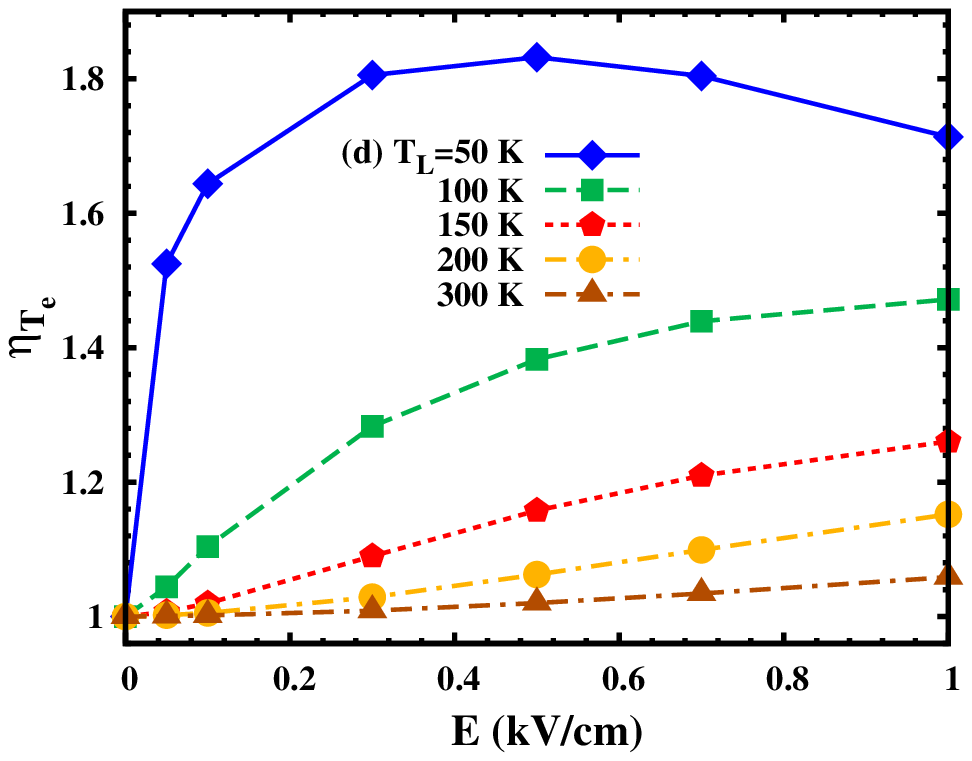}}
      \end{minipage}
      \begin{minipage}[]{10cm}
        \hspace{-1.5 cm}\parbox[t]{5cm}{
          \includegraphics[width=4.5cm,height=4.5 cm]{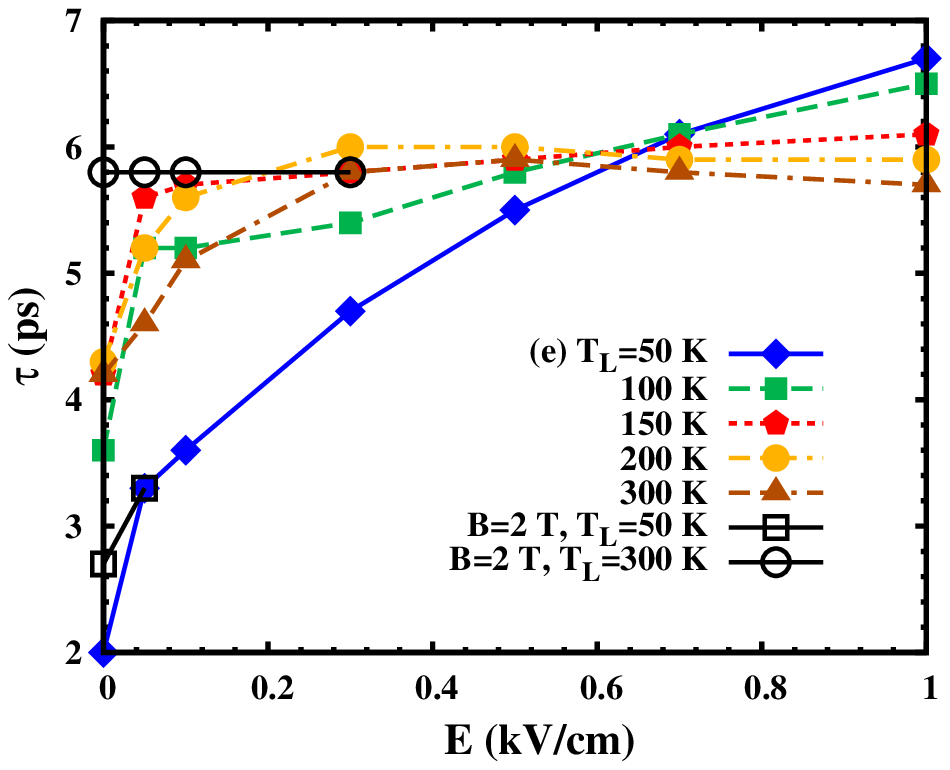}}
        \hspace{-0.7 cm}\parbox[t]{5cm}{
          \includegraphics[width=4.5cm,height=4.5 cm]{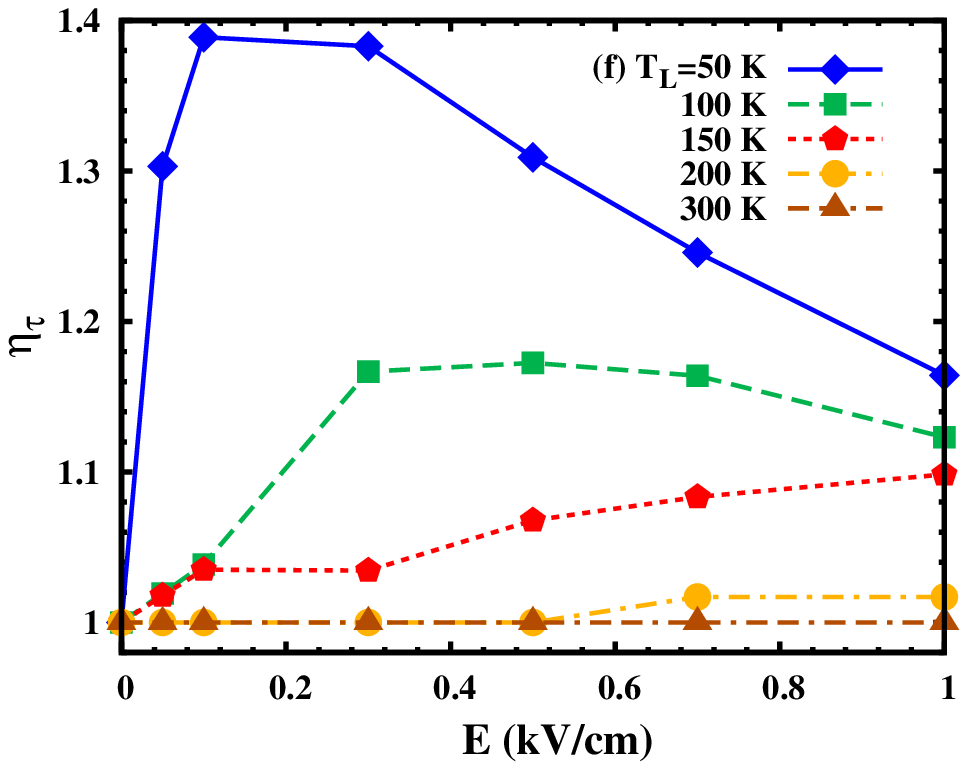}}
      \end{minipage}
     \caption{(Color online) Electric field dependences of drift
       velocity $v_x$ (a), hot-electron temperature $T_e$ (c) and spin relaxation
       time $\tau$ (e), calculated with the equilibrium LO
       phonons.  The ratios of the
       quantities $v_x$, $T_e$ and $\tau$ with the nonequilibrium LO phonons to
       those with the equilibrium ones, $\eta_{v_x}$, $\eta_{T_e}$ and $\eta_\tau$,
       are shown in (b), (d) and (f), respectively.
In (e), the two curves with open squares and open
  circles are the electric field dependences of $\tau$ with a
       magnetic field ($B=2$~T) along the $-{\bf\hat{x}}$-direction
 for $T_L=50$~K and 300~K, respectively [at $T_L=50$~K
       ($T_L=300$~K), when $E>0.05$~kV/cm ($E>0.3$~kV/cm)  
     the spin relaxation time coincides with
     that without magnetic field  and thus not shown].}
\label{figzw2}
\end{figure}
    
In Fig.~\ref{figzw2}(a) the electric field dependence of electron drift
velocity $v_x$ at various lattice temperatures is plotted, with the LO
phonons treated as the equilibrium ones in the calculation. The ratio of $v_x$
obtained with the nonequilibrium LO phonons to that with the equilibrium ones
is shown in Fig.~\ref{figzw2}(b). From Fig.~\ref{figzw2}(b), one notices
that when the nonequilibrium phonon effect is taken into account, $v_x$ can be either increased or
decreased. In fact, the influence of nonequilibrium phonons on 
hot-electron transport consists of two competing effects: the
reabsorption of momentum from the
nonequilibrium phonons tends to increase the electron mobility, while the
enhanced electron heating strengthens the electron-phonon scattering
(including both the electron-AC phonon and electron-LO
phonon scatterings) and tends to  decrease the electron
mobility.\cite{mick193312,cai2636,vai9886} Generally the former
(latter) effect dominates in the regime with low (high) electric filed and/or high (low)
temperature,\cite{mick193312,cai2636,vai9886} as indicated in Fig.~\ref{figzw2}(b). As mentioned
previously, when the electric field is low and/or the lattice
temperature is high, the valley of the LO phonon distribution in $q_x$-negative momentum
space is pronounced, thus it substantially suppresses the back scattering
 of electrons in momentum space by 
absorbing the $q_x$-negative phonons and hence makes the electron distribution more
forward-peaked. Finally, when the lattice temperature is high enough (e.g.,
$T_L=300$~K), the effect of nonequilibrium phonons becomes weak and less
sensitive to the electric field, mainly due to the shorter LO phonon
relaxation time ($\tau_{pp}\approx 1.9$~ps when $T_L=300$~K).\cite{vallee12049}

From Fig.~\ref{figzw2}(c), where the electric field dependence of the 
hot-electron temperature $T_e$ under various lattice temperatures is shown,
one indeed finds that the heating of electrons by the electric field is
quite obvious when the electric field is high and the lattice
temperature is low.\cite{weng1,zhang,jiang} From Fig.~\ref{figzw2}(d), where the ratio of 
$T_e$ obtained with the nonequilibrium LO phonons
to that with the equilibrium ones is plotted, one finds that with the
nonequilibrium phonon effect considered, electrons are further
heated as expected.\cite{cai2636,vai9886,vai13082} Moreover, when $T_L$ is low, $\eta_{T_e}$ shows a
nonmonotonic behavior. That is caused by the decay of the
heating efficiency with the increase of electric field in the presence of nonequilibrium
phonons: With the increase of electric field, the number of the LO phonons
increases and the rate of electron energy relaxation through the 
electron-LO phonon scattering increases as well. This effect is more 
pronounced when the lattice temperature is low, where the
nonequilibrium LO phonons can be considerably accumulated with the
increase of electric field, due to the long phonon relaxation time
($\tau_{pp}=7.3$~ps when $T_L=50$~K)\cite{vallee12049} as well as the
small equilibrium phonon distribution.

The electric field dependence of the spin relaxation time $\tau$ under various
temperatures is plotted in Fig.~\ref{figzw2}(e). From
the figure, one notices that $\tau$
generally increases with $E$ in the regime under
investigation. Two reasons lead to this
phenomenon: (I) Under the electric
field along the $-{\bf\hat{x}}$-direction, a net effective magnetic
field along the $-{\bf\hat{x}}$-direction is induced via the Dresselhaus spin-orbit
coupling.\cite{weng1,zhang} With this effective magnetic field, spins begin to
precess around it and thus the in-plane spin relaxation is 
mixed with the out-of-plane one.\cite{jiang155201,dohr147405} The two-dimensional 
electron system in (001) GaAs quantum well has an in-plane spin relaxation rate smaller than the
out-of-plane one in the framework of the D'yakonov-Perel' relaxation
mechanism. Thus when the electric field is applied, $\tau$
is increased due to the effective magnetic
field.\cite{jiang155201,dohr147405} The effective magnetic field
decreases with the increase of $T_L$, because it is proportional to
$v_x$\cite{weng1,zhang} and $v_x$ decreases with increasing $T_L$ [as
shown in Fig.~\ref{figzw2}(a)]. Therefore  
the mixing of the in-plane and out-of-plane spin relaxations is obvious in the low temperature
regime. In fact, when $E=0.05$~kV/cm, the effective magnetic
field is $\sim$ 1~T when $T_L\sim$50-100~K and $\sim 0.1$~T when $T_L=300$~K. As a result, the effective magnetic field causes an abrupt increase of
$\tau$ [Ref.~\onlinecite{jiang155201}] with the increase of $E$ from 0 to 0.05~kV/cm for the cases
with $T_L\sim$50-100~K but a slow increase of
$\tau$ with the increase of $E$ from 0 to 0.3~kV/cm for the case
with $T_L=300$~K. (II) The heating of electrons by the electric field enhances the
electron-phonon scattering and thus increases the spin relaxation time
$\tau$ in the strong scattering
limit.\cite{weng1,weng2,zhang,zhou,jiang,wu-rev} This effect, only important in the low temperature regime where 
the heating effect is strong [as shown in Fig.~\ref{figzw2}(c)],
is responsible for the continuing increase of $\tau$ with $E$ when
$E>0.05$~kV/cm and $T_L\sim$50-100~K. To make the
underlying physics depicted above more pronounced, a magnetic field $B=2$~T is applied along the
$-{\bf\hat{x}}$-direction for the cases with $T_L=50$~K and 300~K. The
corresponding electric field dependences of $\tau$ are plotted as
curves with open squares ($T_L=50$~K) and open circles ($T_L=300$~K)
in Fig.~\ref{figzw2}(e). With this large external magnetic field, the
in-plane and out-of-plane spin relaxations are efficiently mixed even when
$E=0$. One finds that for the case with $T_L=300$~K, $\tau$ almost
keeps unchanged with the increase of electric field, while for the case with $T_L=50$~K $\tau$ keeps on
increasing with $E$ due to the strong heating effect. Finally, it is noted
that when $T_L=300$~K, there is a marginally decreasing tendency 
of $\tau$ with $E$ when $E$ is near 1~kV/cm. This is caused by the enhanced inhomogeneous
broadening of the effective magnetic field from the Dresselhaus
spin-orbit coupling due to the drifting and heating of the electric
field.\cite{weng1,weng2,zhang,zhou,jiang,wu-rev} This effect is easier to 
take place when the lattice temperature is high.\cite{weng1,zhang}

\begin{figure}[ht]
      \begin{minipage}[]{10cm}
        \hspace{-1.5 cm}\parbox[t]{5cm}{
          \includegraphics[width=4.5cm,height=4.5 cm]{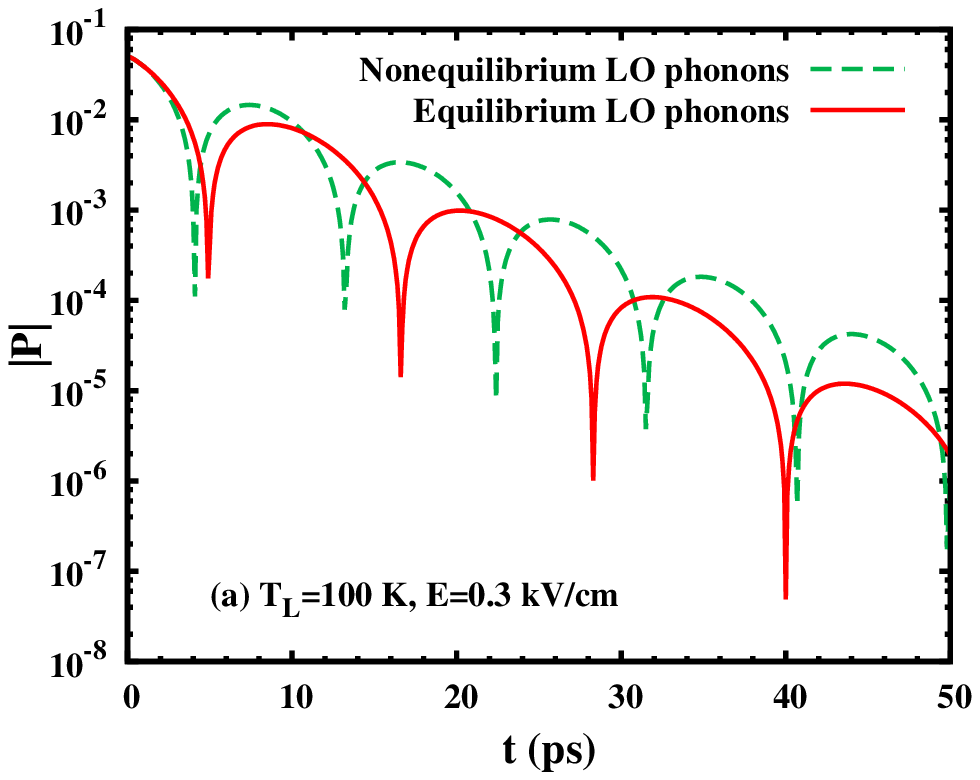}}
        \hspace{-0.7 cm}\parbox[t]{5cm}{
          \includegraphics[width=4.5cm,height=4.5 cm]{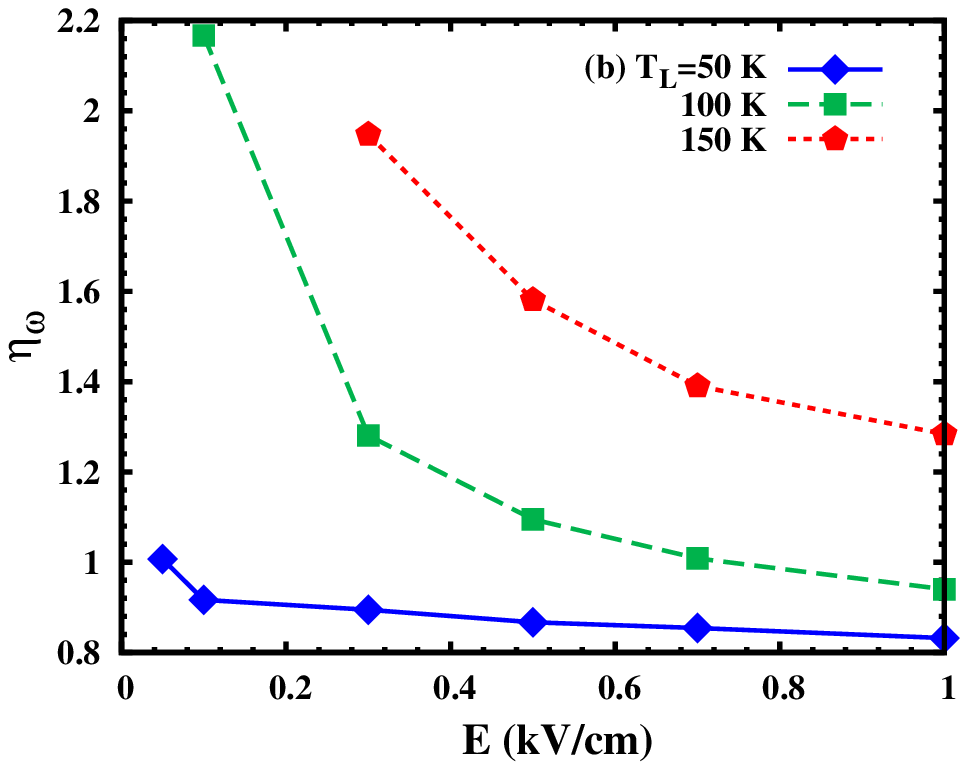}}
      \end{minipage}
      \caption{(Color online) (a) Typical spin precession signals
        calculated with the equilibrium and
        nonequilibrium LO phonons. Note that the
        absolute value of spin polarization $|P|$ is in the logscale. $T_L=100$~K and
        $E=0.3$~kV/cm. (b) The ratio of spin precession frequency
        obtained with the nonequilibrium LO phonons to that with the equilibrium ones, shown
        at three typical temperatures in the electric field region
        where the spin precession signal can be clearly distinguished. }
\label{figzw3}
    \end{figure}

In Fig.~\ref{figzw2}(f), the ratio of the spin relaxation time obtained with the
nonequilibrium LO phonons to that with the equilibrium ones is  
shown. With the nonequilibrium LO phonons taken into account, $\tau$ is
generally increased due to the strengthened electron-phonon
scattering. Therefore the increase of $\tau$ corresponds to the increase of
$T_e$, as shown in Fig.~\ref{figzw2}(d). However, when the lattice temperature
is high enough (e.g., $T_L=300$~K), the modification on $\tau$ induced
by the nonequilibrium phonons can not be seen. Moreover, as the spin precession frequency is 
proportional to $v_x$
(Refs.~\onlinecite{weng1} and \onlinecite{zhang}) and $v_x$ is
affected largely by the nonequilibrium LO phonons, the spin precession
frequency has  a modification with the magnitude roughly
proportional to the change of $v_x$. In Fig.~\ref{figzw3}(a) we show the
typical spin precession signals with the equilibrium and
nonequilibrium  LO phonons, respectively. In
Fig.~\ref{figzw3}(b), the typical ratio of spin precession frequency
obtained with the 
nonequilibrium LO phonons to that with the equilibrium ones is plotted in
the region of the electric field where $v_x$ is large enough that spin precession signal with clear
periods can be distinguished.

Finally, further calculations show that the effect of nonequilibrium
phonons on electron mobility, electron heating and electron spin
relaxation decays with the decrease of electron density and the increase
of impurity density. This can be easily understood as the
LO phonons are driven to the nonequilibrium states by electrons and the
increase of the electron-impurity scattering suppresses the effect
caused by the electron-LO phonon scattering.

\section{Couclusion}
In this work, we have studied the effect of nonequilibrium LO phonons on
hot-electron spin relaxation in $n$-type (001) GaAs quantum
wells. Under an in-plane electric field, the LO phonons can be driven away from their equilibrium states by
electrons and then in turn affect the electron
transport, electron heating as well as electron spin relaxation. 

In the presence of the nonequilibrium LO phonons, the electron drift velocity under
electric field can be either increased (at low electric field and/or
high lattice temperature) or decreased (at high electric field
and/or low lattice temperature). This
phenomenon is caused by the two competing effects: the momentum
reabsorption from phonons and the strengthened electron-phonon
scattering.\cite{mick193312,cai2636,vai9886} The former tends to
increase the electron mobility whereas the latter tends to suppress it. The nonequilibrium LO phonons 
also impede the energy relaxation of electrons and thus the electrons
are further heated, especially when the lattice temperature is
low. The nonequilibrium LO phonons effectively affect the hot-electron
spin relaxation through the strengthening of the electron-phonon scattering,
which tends to increase the spin relaxation time. This effect also dominates
in the low temperature regime. Moreover, as the spin precession
frequency under the electric field is proportional to the electron drift velocity, it can be either
increased or decreased when the nonequilibrium LO phonons are taken into
account. Finally, it should be noticed that the effect of nonequilibrium phonons
is more pronounced in systems with high electron density and low
impurity density. 

\begin{acknowledgments}
This work was supported by the Natural Science Foundation of China
under Grant No.~10725417, the
National Basic Research Program of China under Grant 
No.~2006CB922005 and the Knowledge Innovation Project of Chinese Academy
of Sciences. 
\end{acknowledgments}

\end{document}